\documentclass[11pt,a4paper]{article}

\pdfoutput=1

\usepackage{jheppub}
\usepackage{amsmath}
\usepackage{amsfonts}
\usepackage{graphicx}
\usepackage{amssymb}
\usepackage{amssymb}
\usepackage{latexsym}
\usepackage{array}
\usepackage{bbold}
\usepackage{comment}
\usepackage{cancel}
\usepackage{ulem}

\newcommand{\bea}{\begin{eqnarray}}
\newcommand{\eea}{\end{eqnarray}}
\newcommand{\be}{\begin{equation}}
\newcommand{\ee}{\end{equation}}

\newcommand{\re}[1]{(\ref{#1})}

\usepackage{xcolor}
\definecolor{mygreen}{HTML}{006E28}

\newcommand{\delete}[1]{}
\title{Perturbations of Q-balls: from spectral structure to radiation pressure}

\author[a,d]{Dominik Ciurla,}
\author[b]{Patrick Dorey,}
\author[a]{Tomasz Roma\'nczukiewicz}
\author[c]{and Yakov Shnir}

\affiliation[a]{Faculty of Physics, Astronomy and Applied Computer Science, Jagiellonian University, Krak\'ow, Poland}
\affiliation[b]{Department of Mathematical Sciences, Durham University, Durham DH1 3LE, UK}
\affiliation[c]{Institute of Physics, University of Oldenburg,
Oldenburg D-26111, Germany\\
Hanse-Wissenschaftskolleg, Lehmkuhlenbusch 4, 27733 Delmenhorst, Germany}
\affiliation[d]{Doctoral School of Exact and Natural Sciences, Jagiellonian University, Krak\'ow, Poland}

\emailAdd{dominik.ciurla@doctoral.uj.edu.pl}
\emailAdd{p.e.dorey@durham.ac.uk}
\emailAdd{tomasz.romanczukiewicz@uj.edu.pl}
\emailAdd{shnir@maths.tcd.ie}

\abstract{We investigate Q-balls in a 1+1 dimensional complex scalar field theory. We find that the relaxation of a squashed Q-ball is dominated by the decay of a normal mode through nonlinear coupling to scattering modes and a long-lasting quasi-normal 
mode. We also study how these Q-balls behave when exposed to scalar radiation, finding that for certain conditions they can experience negative radiation pressure.}

\begin{document}

\maketitle

\section{Introduction}
Q-balls are non-topological solitons which arise in various non-linear scalar field theories
possessing an unbroken continuous global symmetry \cite{Rosen:1968mfz,Friedberg:1976me,Coleman:1985ki}.
They carry a Noether charge associated with this symmetry, and are stationary
localized field configurations with an explicitly time-dependent phase.
Typical examples of Q-balls in flat 3+1 dimensional Minkowski spacetime are spherically symmetric
solutions of a model with a single complex scalar field and
a suitable  self-interaction potential \cite{Coleman:1985ki}, or solitons of
the renormalizable Friedberg-Lee-Sirlin two-component model with a
symmetry breaking potential \cite{Friedberg:1976me}. Physically, they can be considered as a condensate of a large number of scalar quanta which, for a fixed value of the charge $Q$,
yields an extremum of the effective energy functional.
In such a context, the charge can also be interpreted as the particle number.

Q-balls have received considerable attention over the last
three decades (for a review, see, e.g., \cite{Lee:1991ax,Radu:2008pp,shnir_2018}). However, most works address the stability of stationary Q-balls
and  the domains of their existence. There have only been a few studies of the dynamics of Q-balls
\cite{Axenides:1999hs,Battye:2000qj,Bowcock,Copeland:2014qra},
which appears to be very different from the usual dynamics of topological solitons. The reasons for this are
related to phase-dependent force of interaction between the Q-balls,
their non-topological charge which can be transferred in collisions
\cite{Copeland:2014qra}, and the complicated spectrum of their excitations,
which as we will show below may include both normal and quasinormal (QNM) modes (for  discussions of QNM see \cite{chandrasekhar1975quasi,Kokkotas:1999bd,Konoplya:2011qq} in the context of black holes, \cite{PhysRevLett.92.151802} in relation with excitations of a non-Abelian monopole, 
and~\cite{ching1998quasinormal}). Furthermore, in some cases these excitations cannot be considered
as linearized perturbations, since at least second order corrections must be taken into account in order to understand them fully \cite{Smolyakov1}.

There is some similarity between Q-balls and oscillons \cite{Bogolyubsky:1976nx,Copeland:1995fq,Gleiser:1993pt},
extremely long-lived, spatially localized, almost periodic non-linear field configurations. It was pointed out that a Q-ball can be roughly viewed as a system of two interacting oscillons, associated with
real components of a complex scalar field
\cite{Copeland:2014qra}. However,
while Q-balls are non-radiating stationary solutions,  oscillons  slowly radiate  energy 
\cite{Fodor:2006zs,Fodor:2008du,Grandclement:2011wz}. A peculiar feature of the radiation of an oscillon is its  resonant
character \cite{Honda:2001xg,Dorey:2019uap,Zhang:2020bec}: the radiating oscillon  can pass through a sequence of quasi-stable
Q-ball-like configurations. The relation between Q-balls and oscillons is supported by the existence
of a so-called adiabatic invariant that is approximately conserved \cite{Kasuya:2002zs,Kawasaki:2015vga,Levkov:2022egq}.

Important information about the properties of solitons can be obtained from the study of their interactions with
external perturbations. In particular, one can
consider a small amplitude incoming wave moving
towards a soliton. It has been observed that in a large class of models with kink solutions the radiation pressure exerted on the kink can be negative \cite{Romanczukiewicz:2003tn,Romanczukiewicz:2008hi,Forgacs:2008az}. Similar effects were found in Bose-Einstein condensates \cite{NRP_IN_BE}.
Scalar radiation may also strongly influence the dynamics of soliton collisions \cite{Dorey:2023izf}, so a natural question arises as to what the effect of the interaction of Q-balls with incoming scalar radiation might be. 

In simple cases in (1+1) dimensions, analytical solutions for stationary Q-balls are known, as discussed, for example, in \cite{Bowcock}. 
The goal of this paper is to study perturbative excitations of some of these Q-balls. 
In particular, we reconsider squashing perturbations of a single Q-ball \cite{Bowcock} and examine the
effects of radiation pressure.
The paper is organized as follows. In section \ref{sec:model}, we define the model and some important properties of the Q-balls. Section \ref{sec:spectral_structure} is devoted to the linearization problem and spectral structure of the Q-balls. We analyze scattering and bound modes, as well as the so-called half-propagating modes, investigating in particular their connection with quasinormal modes. We analyze how the spectral structure, and especially the bound and quasinormal modes, influences the long-term evolution of a perturbed Q-ball. In section \ref{sec:radiationpressure}, we focus on the motion of Q-balls interacting with a monochromatic wave. We find that for a certain set of parameters, it is possible for Q-balls to accelerate towards the source of radiation, which is an indication of the negative radiation pressure. After our conclusions in section \ref{sec:conclusions}, we comment in appendix \ref{metaqballs} on the results of \cite{Bowcock} for a metastable Q-ball.

\section{A class of Q-balls in 1+1 dimensions}\label{sec:model}
\subsection{The model}
\begin{figure}
    \centering
     \includegraphics[width=0.75\textwidth]{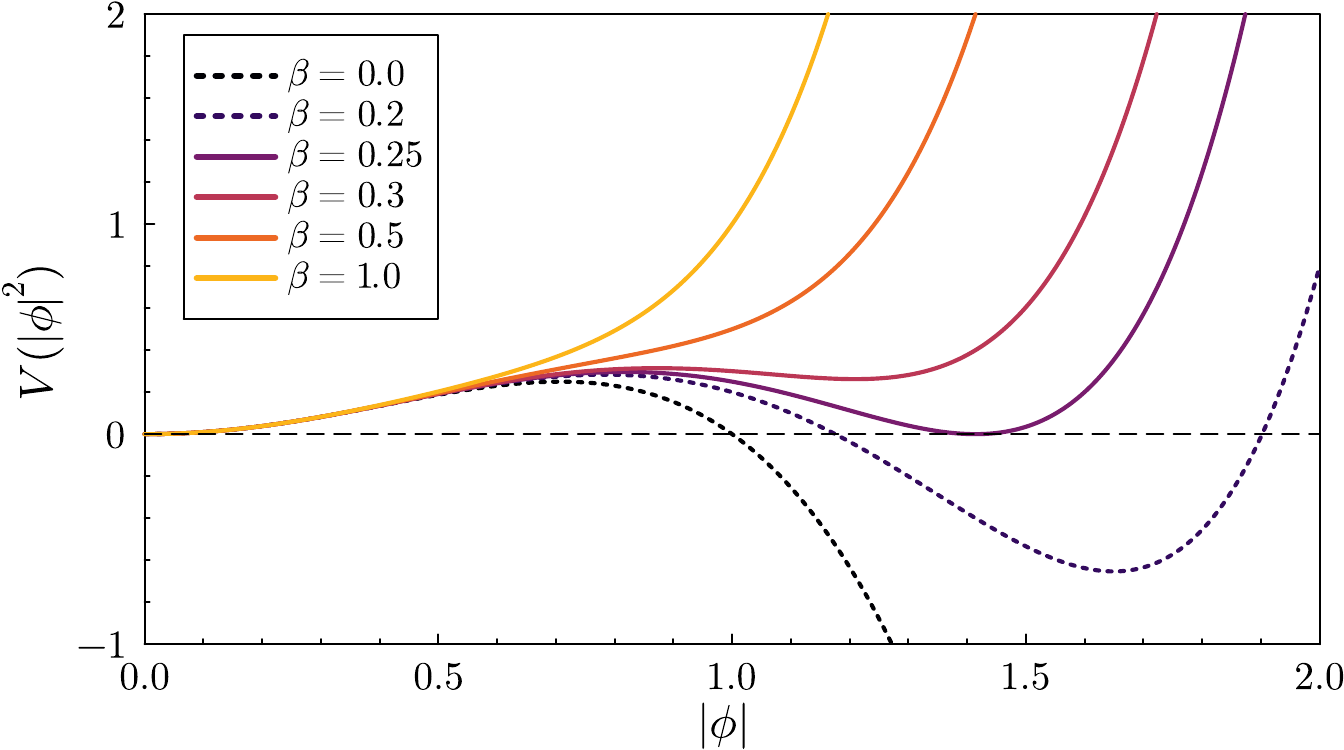}
     \caption{The field theory potential $V(|\phi|^2)$ defined by (\ref{pot}) for various values of $\beta$. For the potentials shown with dotted lines, the vacuum at $|\phi|=0$ is metastable.}
     \label{fig:theory_potentials}
\end{figure}

The (1+1)-dimensional scalar field theory we will study is defined by the following Lagrangian \cite{Axenides:1999hs,Bowcock}:
\begin{equation}
    \mathcal{L} = \partial_\mu\phi\partial^\mu\phi^*-V\left(|\phi|^2\right)\, ,
    \label{lag}
\end{equation}
where the asterisk denotes complex conjugation and the rescaled potential of the self-interacting complex scalar field $\phi$ is
\begin{equation}
V\left(|\phi|^2\right)  =|\phi|^2-|\phi|^4+\beta |\phi|^6\,.
\label{pot}
\end{equation}
This potential is shown for various values of the parameter $\beta$ in figure \ref{fig:theory_potentials}. There is a minimum at $\phi=0$ which for $\beta>1/4$ is the global minimum, while
for $\beta<1/3$ there is a second minimum  at 
$|\phi|^2=1+\sqrt{1-3\beta}$\,.

The global $U(1)$ symmetry of the Lagrangian (\ref{lag}) corresponds to the conserved current
\begin{equation}
j_\mu=i(\phi^* \partial_\mu \phi - \partial_\mu \phi^* \phi)\, , \quad \partial^\mu j_\mu=0\, .
\label{charge}
\end{equation}

Variation of the Lagrangian (\ref{lag}) with respect to the scalar field leads to the  equation of motion
\begin{equation}
    \phi_{tt}-\phi_{xx}+\frac{\partial V}{\partial |\phi|^2}\phi=0
\label{field_eq}
\end{equation}

A stationary (1+1)-dimensional Q-ball configuration with harmonic time dependence can be parameterised as
\begin{equation}
    \phi(x, t) = e^{i\omega t}f(x)
\label{ansatz}
\end{equation}
where $f(x)$ is a real profile function which satisfies the  first order equation
\begin{equation}
\label{stationary}
\frac{df}{dx}=\pm\sqrt{\tilde V(f^2)}
\end{equation}
with the boundary conditions $f(\pm\infty)=0$, and
\begin{equation} \label{effective_pot}
    \tilde V(f^2) = V(f^2) -\omega^2 f^2
\end{equation}
is the effective potential for the Q-ball configuration. Further specifying $f_x(0)=0$ centres the profile function on the origin.

The Noether charge of the stationary configuration is
\be
\label{Q-charge}
Q= \int dx j_0 = 2 \omega \int dx f^2 = 2 \omega N \, ,
\ee
where $N$ is the $L^2$ norm of the scalar field.
The total energy of the Q-ball can be written as
\be
\label{energy}
E=\int dx \left\{\left(  f_x\right)^2 +
\omega^2 f^2 + V(f^2) \right\} \, = 
\int dx \left\{\left(  f_x\right)^2 +
\tilde V(f^2) \right\} + \omega Q\, .
\ee

\subsection{Stationary solutions}
Equation (\ref{stationary}) admits the analytic solution
\begin{equation}
    f(x;\omega) = \frac{\sqrt{2}\omega'}{\sqrt{1+\sqrt{1-4\beta\omega'^2}\cosh(2\omega'x)}}\, ,
\label{solution}
\end{equation}
where $\omega'=\sqrt{1-\omega^2}$ is the complementary frequency \cite{Bowcock}. For the sake of simplicity, we will drop $\omega$ and write a shorthand version $f(x)$ in most cases, except for one paragraph in section \ref{sec:nonoscillating}.

Localized soliton solutions of the model (\ref{lag}) exist only for a limited range of values of the angular frequency, $\omega_{\rm min}\le \omega \le \omega_{\rm max}$. Here the upper bound corresponds to mass of the linearized scalar excitations, $\omega_{\rm max} = V^{\prime \prime}(0)/2=1$, and the lower bound depends on the value of the parameter $ \beta$ as \cite{Bowcock}
\begin{equation}
    \omega_{\rm min}(\beta) = \sqrt{1-\frac{1}{4\beta}}\, .
    \label{wmin}
\end{equation}

Note that the potential (\ref{pot}) possesses a
unique global minimum at $\phi=0$ if
$\beta > \frac{1}{4}$. If $\beta < \frac{1}{4}$, the local minimum at $\phi=0$ becomes a false vacuum. In such a case, the configuration (\ref{solution}) 
will be metastable.
Finally, in the marginal case $\beta=\frac{1}{4}$ the vacuum $V\left(|\phi|^2\right)=0$ is two-fold degenerate between $\phi_0=0$ and $|\phi_1|=\sqrt 2$ and the model
supports topological solitons, or kinks. In this limit, the minimal value of the angular frequency is zero.

More generally, as the angular frequency approaches the minimal value $\omega_\textrm{min}$ (\ref{wmin}), the effective potential $\tilde V(f^2)$ takes the form of the standard $\phi^6$ potential
\begin{equation}
    \tilde V(f^2) = \beta f^2\left(f^2-\frac{1}{2\beta}\right)^2
\end{equation}
with minima at $\tilde V(0)=\tilde V\left(\frac{1}{2\beta}\right)=0$ and the Q-ball splits into a pair of kink-like solutions interpolating between these vacua \cite{Lohe:1979mh}
\begin{equation}
    \phi_K(x,t) = \frac{e^{i\omega_{\rm min} t}}{2\sqrt{\beta}} \sqrt{1\pm\tanh\left(\frac{x}{2\sqrt{\beta}}\right)}.
\end{equation}
The energy of a kink is
$
E_K = \frac{1}{8}\beta^{-3/2}
$.

Substitution of the ansatz  (\ref{solution})
into  expressions (\ref{Q-charge}) and (\ref{energy}) for the energy and
the charge of the configuration  gives  \cite{Bowcock}
\be
E = \frac{4\omega {\omega'} + Q (4\beta-1 + 4 \beta \omega^2 )}{8\omega \beta}\,  ;\qquad
Q = \frac{4\omega}{\sqrt \beta}{\rm arctanh}\left(\frac{1-\sqrt{1-4\beta {\omega'}^2}}{2{\omega'} \sqrt \beta}\right) \, .
\label{EQ}
\ee

Note,  it is possible to consider Q-balls for $\beta<\frac{1}{4}$, but in such cases they are excitations of a false vacuum and are only metastable solutions. Large perturbations can lead to a collapse to a true vacuum. In \cite{Bowcock} the authors considered a relaxation process of a weakly perturbed Q-ball for $\beta=0$. In this paper we will mostly assume that $\beta>\frac{1}{4}$, although some results for $\beta=0$ will be discussed in the Appendix.

\section{Spectral structure}\label{sec:spectral_structure}
\subsection{Linearization}
\begin{figure}
    \centering
    \includegraphics[width=1\textwidth]{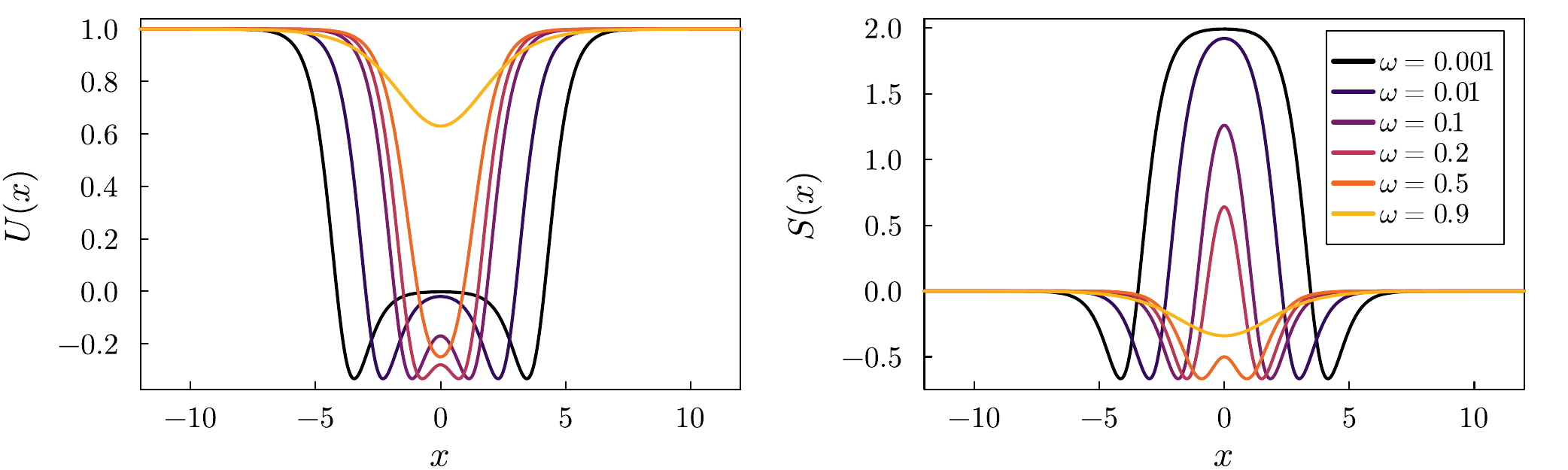}
    \caption{Potentials of linearized perturbations $U(x)$ and $S(x)$, defined in equations (\ref{potentialU}) and (\ref{potentialS}),  for selected values of $\omega$ and $\beta=\frac14$.}\label{fig:potentials}
\end{figure}

Our aim now is to analyse the
spectrum of linearized perturbations of these Q-balls.  By analogy with the analysis of stability of the solutions in \cite{Smolyakov1, Smolyakov2}, we consider a perturbative
expansion of the scalar field
\begin{equation}
    \phi(x, t) = f(x)e^{i\omega t} + A\xi(x,t) + \cdots\,,
\end{equation}
where the real parameter $A$ is the amplitude of the small perturbation of the Q-ball, and substitute
it into the field equation (\ref{field_eq}).
In the linear approximation in $A$
we obtain
\begin{equation}
\ddot \xi - \xi''+\left[V'(f^2)+V''(f^2)f^2\right]\xi + V''(f^2)f^2\xi^*=0 \,.
\label{pert}
\end{equation}
For oscillating modes it is consistent to consider perturbations $\xi$ of the form
\begin{equation}\label{decomposition}
    \xi(x,t) = e^{i(\omega+\rho)t}\eta_1(x) + e^{i(\omega-\rho)t}\eta_2(x) \, ,
\end{equation}
where the parameter $\rho$ encodes the possible eigenfrequencies of the perturbation.
The linearized equation (\ref{pert}) can then
be written as a set of two coupled second order ordinary differential equations for the
components $\eta_1$ and $\eta_2^*$:
\begin{equation}\label{linear}
    L\begin{bmatrix}
        \eta_1\\\eta_2^*
    \end{bmatrix}=0
\end{equation}
\begin{equation}
L=-\begin{bmatrix}
    (\omega+\rho)^2&\\
    &(\omega-\rho)^2
\end{bmatrix} +
\begin{bmatrix}
    D& S\\
    S&D
\end{bmatrix}
\end{equation}
where
\begin{equation}
    D=-\frac{d^2}{dx^2}+U + S
\end{equation}
and  the potentials of perturbations $U(x)$ and $S(x)$ are (see Figure~\ref{fig:potentials})
\begin{equation}\label{potentialU}
    U = V'(f^2)=1-2f^2+3\beta f^4
\end{equation}
\begin{equation}\label{potentialS}
    S = f^2V''(f^2)=f^2(-2+6\beta f^2)
\end{equation}
Note  that $U(x)\to 1$ and $S(x)\to 0$ as $|x|\to \infty$. In these limits the equations for $\eta_1$ and $\eta_2$ decouple.

\begin{figure}
    \centering
    \includegraphics[width=1\textwidth]{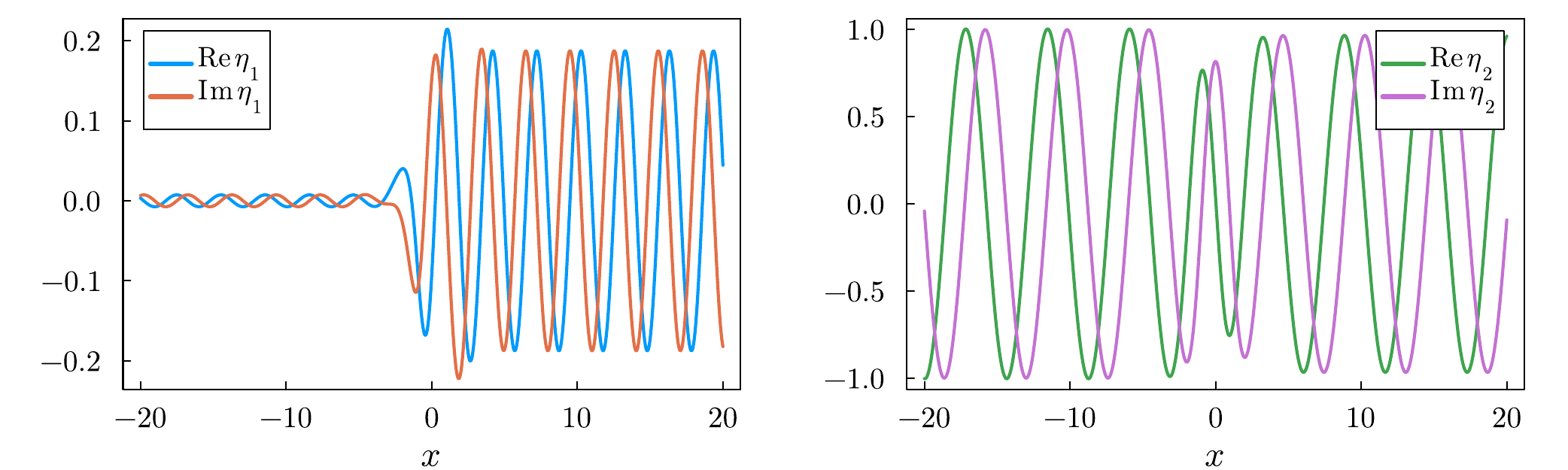}
    \caption{Example solution to the linearized problem for $\beta=\frac{1}{4}$, $\omega=0.8$ and $\rho=2$ with a wave travelling to the right in the second channel. }\label{fig:example_mode}
\end{figure}

\begin{figure}
    \centering
    \includegraphics[width=1\textwidth]{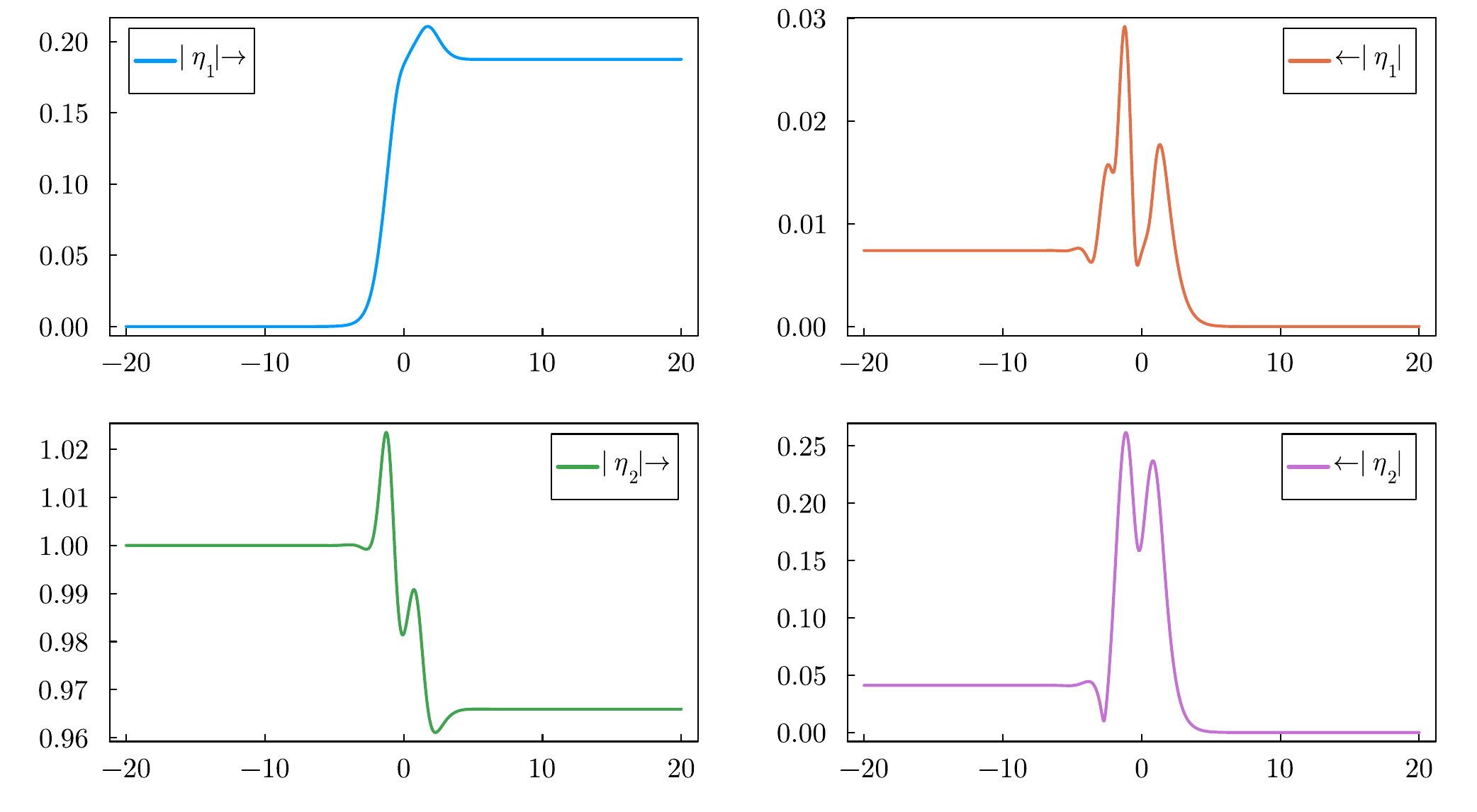}
    \caption{Decomposition of the wave into scattered components for $\beta=\frac{1}{4}$, $\omega=0.8$ and $\rho=2$ with a wave travelling right boundary conditions}\label{fig:example_decomposition}
\end{figure}

Assuming $\omega>0$, we consider the following cases:
\begin{itemize}
\item $\rho\in(0,1-\omega)$: there are no traveling waves, however, bound modes may exist, depending on the form of the potential.
\item $\rho\in (1-\omega, 1+\omega) $: only the component $\eta_1$ is asymptotically propagating, with $e^{-ik_1x}$ the right-moving mode and $e^{ik_1x}$ the left-moving mode, where the wavenumber is $k_1=\sqrt{(\omega+\rho)^2-1}>0$. 
The second component $\eta_2$ remains exponentially localized on the Q-ball, $k_2^2 = (\omega-\rho)^2-1 <0$. 
Throughout this paper we will refer to this state as the half-propagating mode. 
Such modes are crucial for understanding quasinormal modes in other models, including the 't Hooft-Polyakov monopole \cite{PhysRevLett.92.151802}.
\item $\rho > 1+\omega$: both components are propagating, but the component $\eta_2$ oscillates with \textit{negative} frequency, which means that for $k_2>0$ the term $e^{ik_2x}$ describes a wave moving to the right, $k_2=\sqrt{(\omega-\rho)^2-1}>0$. An example of such a mode is presented in figures \ref{fig:example_mode} and \ref{fig:example_decomposition}.
\end{itemize}

\subsection{Zero modes}\label{sec:nonoscillating}
There are two zero modes that satisfy the equation (\ref{linear}) with $\rho=0$.
One of them is the translational mode, which shifts the position of the Q-ball,
\begin{equation}
    \eta_0^{(tr)}(x) = \partial_xf(x); \qquad (f(x)-x_0\eta_0^{(tr)}(x))e^{i\omega t}\approx f(x-x_0)e^{i\omega t} \, .
\end{equation}
The other one is the phase shifting zero mode
\begin{equation}
    \eta_0^{(ph)}(x) = f(x); \qquad (f(x)+i\alpha \eta_0^{(ph)}(x))e^{i\omega t}\approx f(x)e^{i\omega t+i\alpha} \, ,
\end{equation}
which corresponds to the global $U(1)$ symmetry
of the model (\ref{lag}).

However, apart from these two modes, there are two more solutions
of the linearized equation (\ref{pert}), of the form
$\xi(x,t) = \eta(x,t) e^{i\omega t}$ with the $\eta$ factor 
also time dependent:
a stationary mode $\eta^{(Q)}(x,t)$, which transforms the frequency of the Q-ball $\omega$ as  $\omega+\delta \omega$, and a mode corresponding to Lorentz symmetry.
Recall that the profile function $f(x;\omega)$ \re{solution} depends on the frequency, therefore excitation of this mode affects both the frequency and the profile of the Q-ball, thus
\begin{equation}
    \eta^{(Q)}(x,t) = \partial_\omega f(x;\omega)+itf(x;\omega); \qquad (f(x;\omega)+\delta\omega\eta^{(Q)}(x,t))e^{i\omega t}\approx {f}(x;\omega+\delta\omega)e^{i(\omega+\delta\omega) t} \, .
\end{equation}
The Lorentz symmetry mode is
\begin{equation}
    \eta^{(Lor)}(x,t) = t\partial_xf(x)+i\omega x f(x) \, ,
\end{equation}
which follows from the expansion of a boosted Q-ball solution expanded up to $\mathcal{O}(v)$ terms:
\begin{equation}
    f(\gamma(x-vt))e^{i\gamma\omega(t-vx)}\approx \left[f(x)-v\left( t\partial_xf(x)+i\omega x f(x)\right)\right]e^{i\omega t} \, .
\end{equation}
Note that translational, phase shifting and Lorentz zero modes preserve the charge of the Q-ball while the frequency transforming mode changes it.

\subsection{Half-propagating and quasi-normal modes}
For $\rho$ within the range $(1-\omega,1+\omega)$ only one component of the linearized solution propagates, $k_1^2>0$, whereas the other component cannot, $k_2^2<0$. In such a case, the second component $\eta_2$ has to decay as $x\to\pm\infty$, as $e^{\mp|k_2|x}$. 

\begin{figure}
    \centering
    \includegraphics[width=0.75\textwidth]{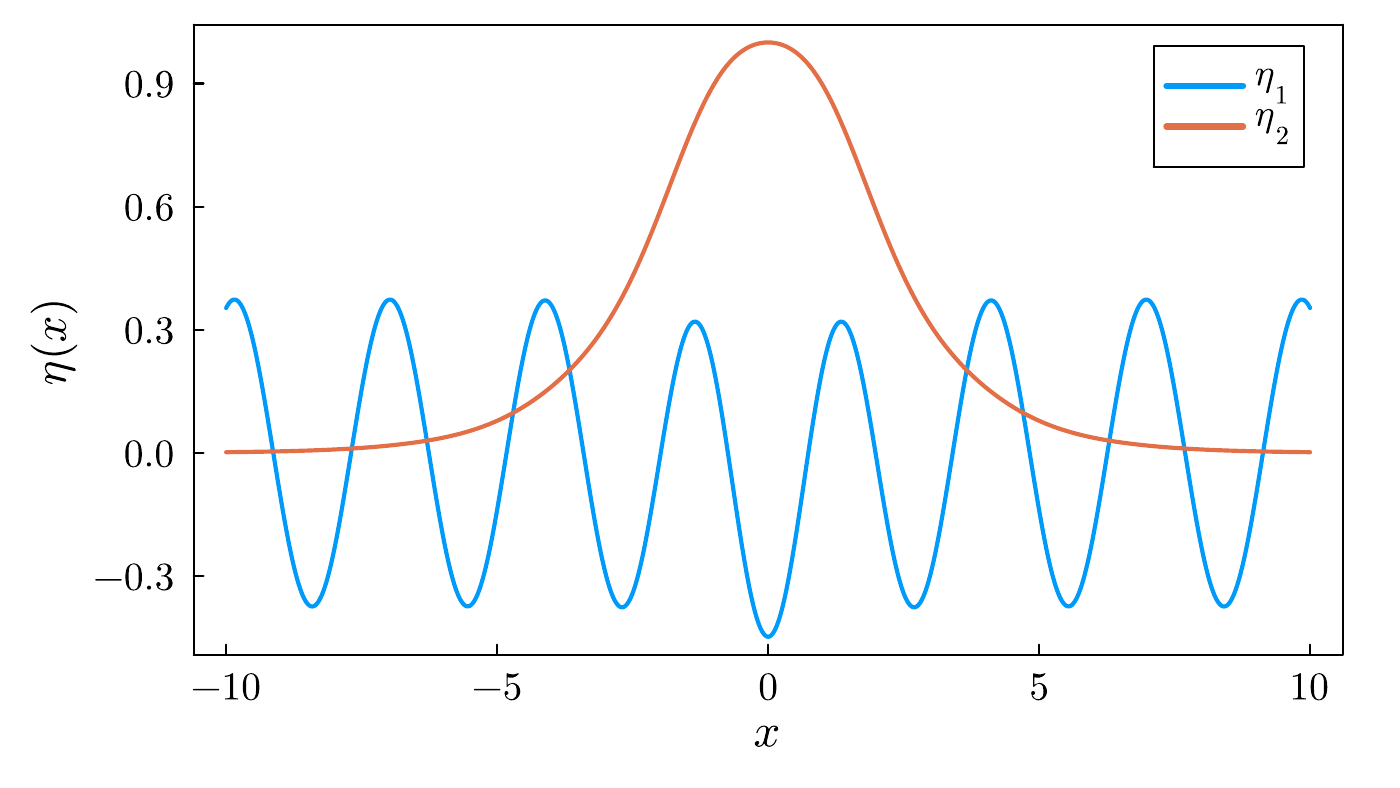}
    \caption{An example of symmetric half-propagating mode for $\beta=\frac{1}{4}$ and $\omega=\frac{\sqrt{3}}{2}$, $\rho=1.54$.}\label{fig:half_bound1}
\end{figure}

In our numerical calculations we fixed 
\begin{subequations}\label{conditions}
\begin{equation}
    \eta_1'(0)=\eta_2'(0)=0
\end{equation}
for a symmetric solution, normalised by setting 
\begin{equation}
    \eta_2(0)=1,
\end{equation}and then adjusted the value of $\eta_1(0)$ so as to satisfy the remaining boundary condition
\begin{equation}
    \eta_2'(x_f)+|k_2|\eta_2(x_f)=0
\end{equation}
\end{subequations}
at some large distance $x_f>0$. As a result of these conditions, far from the Q-ball the profile of the first, propagating, component has the form
\begin{equation}
    \eta_{1}(x)=A_{\rm rad}\cos(k_1|x|+\delta)\,.
\end{equation}
An example of such a symmetric half-propagating solution is presented in figure~\ref{fig:half_bound1}.

\begin{figure}
    \centering
    \includegraphics[width=0.75\textwidth]{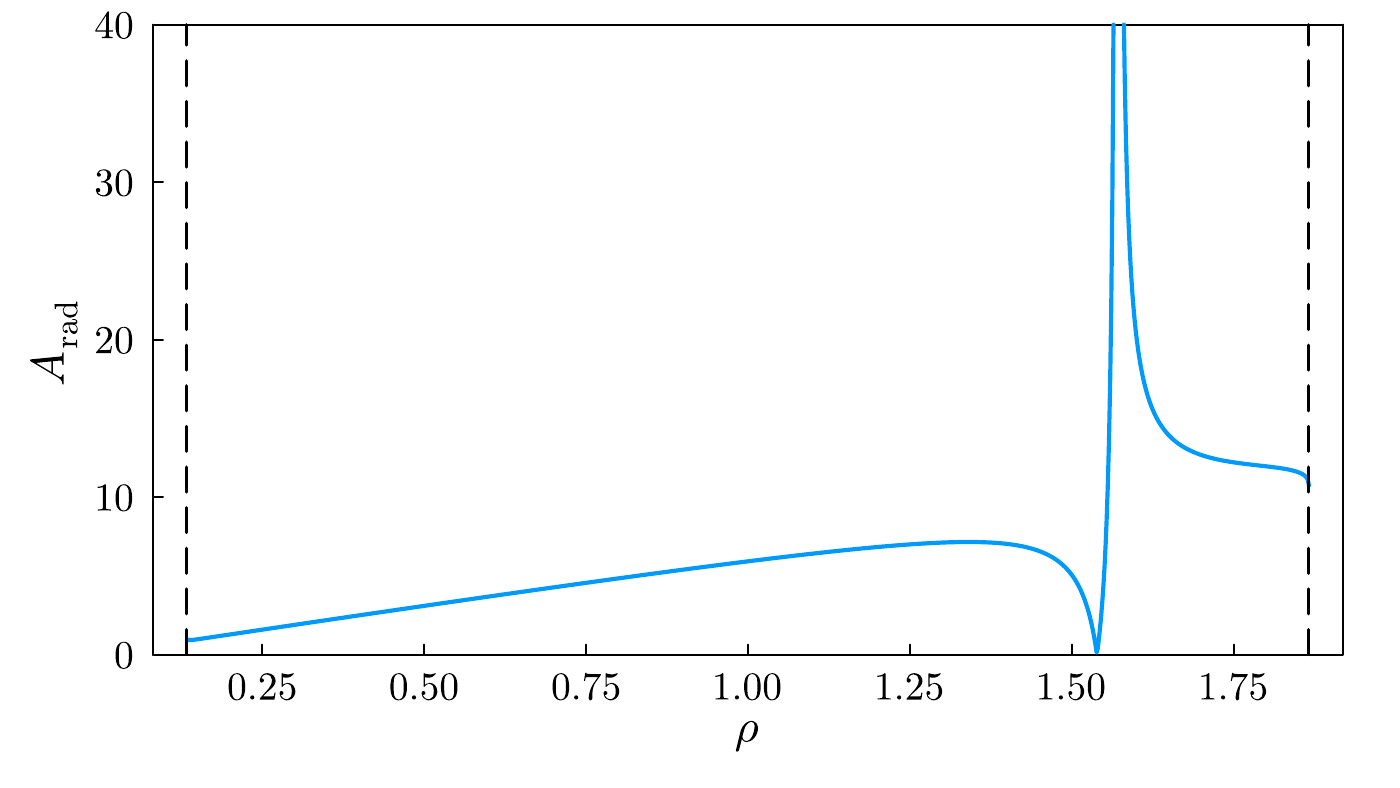}
    \caption{Radiation tails of the symmetric half-propagating modes as a function of $\rho$ for $\beta=\frac{1}{4}$ and $\omega=\frac{\sqrt{3}}{2}$.}\label{fig:half_bound_amplitudes}
\end{figure}

\begin{figure}
    \centering
    \includegraphics[width=0.75\textwidth]{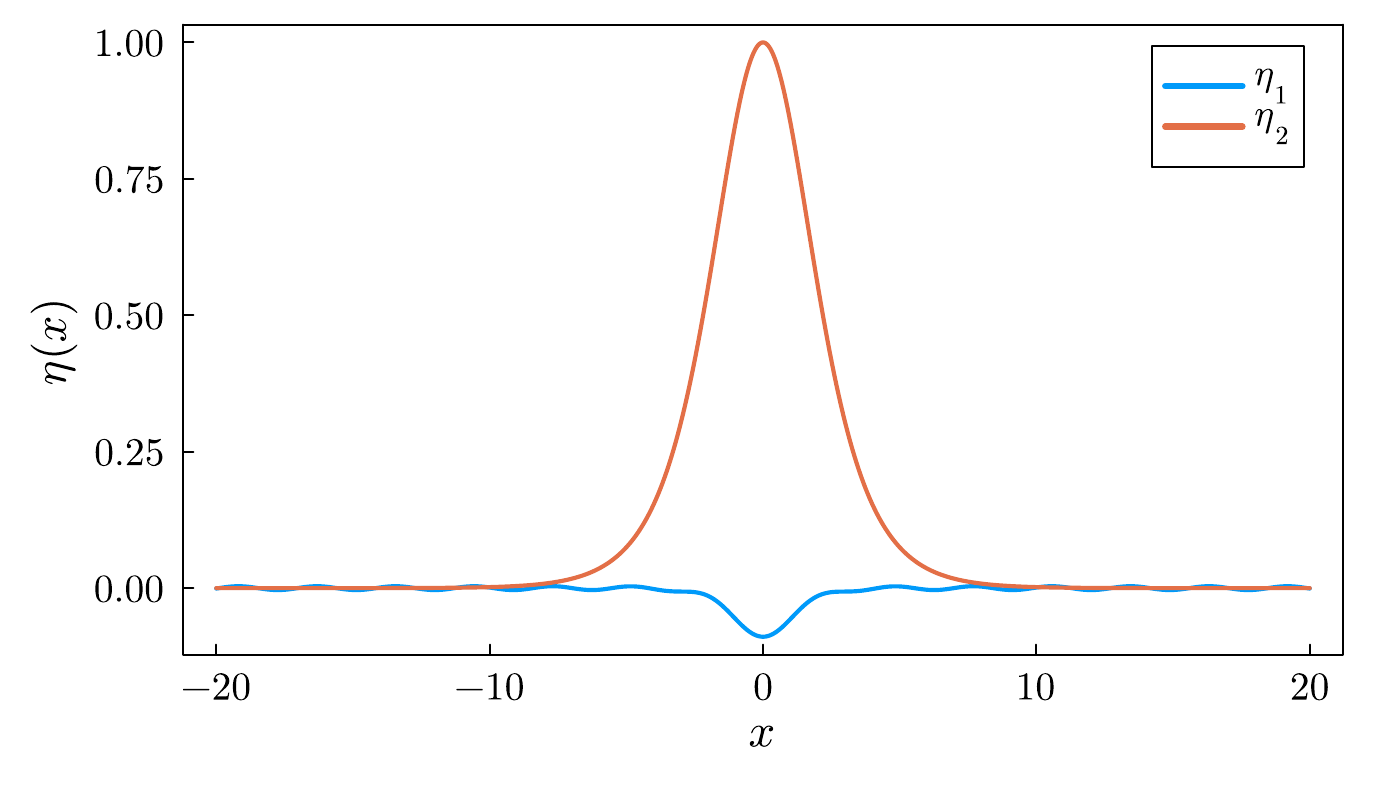}
    \caption{profile of the half-propagating mode with minimal radiation tail for $\beta=\frac{1}{4}$ and $\omega=\frac{\sqrt{3}}{2}$ and $\rho=1.538789$.}\label{fig:half_bound_min_amp}
\end{figure}

The consequences of the excitation of the half-propagating mode are twofold. First, the non-propagating part $\eta_2(x)$ has an impact on the profile of the Q-ball. Second, the propagating component of this mode $\eta_1(x)$ radiates away the energy of the perturbed configuration. We can expect that the modes with the largest amplitude  would decay faster, and conversely, the modes with the smallest amplitude would stay longer. 
In figure~\ref{fig:half_bound_amplitudes} we show the amplitude $A_{\rm rad}$ of the propagating component as a function of  $\rho$ for $\beta=\frac{1}{4}$ and $\omega=\frac{\sqrt{3}}{2}$. 
There are two important features. First, around $\rho=1.57$ this amplitude grows almost to infinity, which indicates that our boundary conditions (\ref{conditions}) cannot be simultaneously fulfilled, which basically means that $\eta_2(0)$ should vanish. Second, before this discontinuity, the amplitude of the propagating mode becomes very small, $A_\textrm{rad}=0.0034979$ for $\rho=1.538789$. 
Such a minimal radiation tail can, and in this case does, indicate the presence of a quasinormal mode \cite{Fodor:2006zs}. Quasinormal modes satisfy purely outgoing boundary conditions, which break the Hermiticity of the operator $L$ in (\ref{linear}). This allows the eigenfrequency to be complex.  And indeed, we have found a quasinormal mode at $\rho=1.538789 + 1.180\cdot10^{-5}i$.
The half-life of this mode is $T_{1/2} = \frac{1}{\textrm{Im}\,\rho}\log(2)\approx 58700$ which explains the longevity of the oscillations. The profile of the half-propagating mode with minimal radiation tail is shown in figure~\ref{fig:half_bound_min_amp}.

\subsection{Bound modes}
For perturbation frequencies $\rho\in(0,1-\omega)$ neither of the components can propagate, and the profiles have only exponential tails. The corresponding normalizable modes are bound to 
the Q-ball. We found various examples of these modes, including\footnote{For most of our analysis we consider stable Q-balls with $\beta\ge\frac14$, but we devote Appendix \ref{metaqballs} to the relaxation problem for $\beta=0$.} 
in the case $\beta=0$ analysed in~\cite{Bowcock}.
As mentioned above, there always is a translational zero mode and one bound mode for $\rho=0.1336$, which is very close to the frequency of one of the peaks in the spectrum.  The component for $\omega-\rho$ is localized at $x=0$, the other $\omega+\rho$ is very close to the threshold, but $\eta_2(0)$ is much smaller than $\eta_1(0)$, which explains why there is no peak near the threshold in the power spectrum of $\phi(0, t)$.

\begin{figure}
    \centering
    \includegraphics[width=1\textwidth]{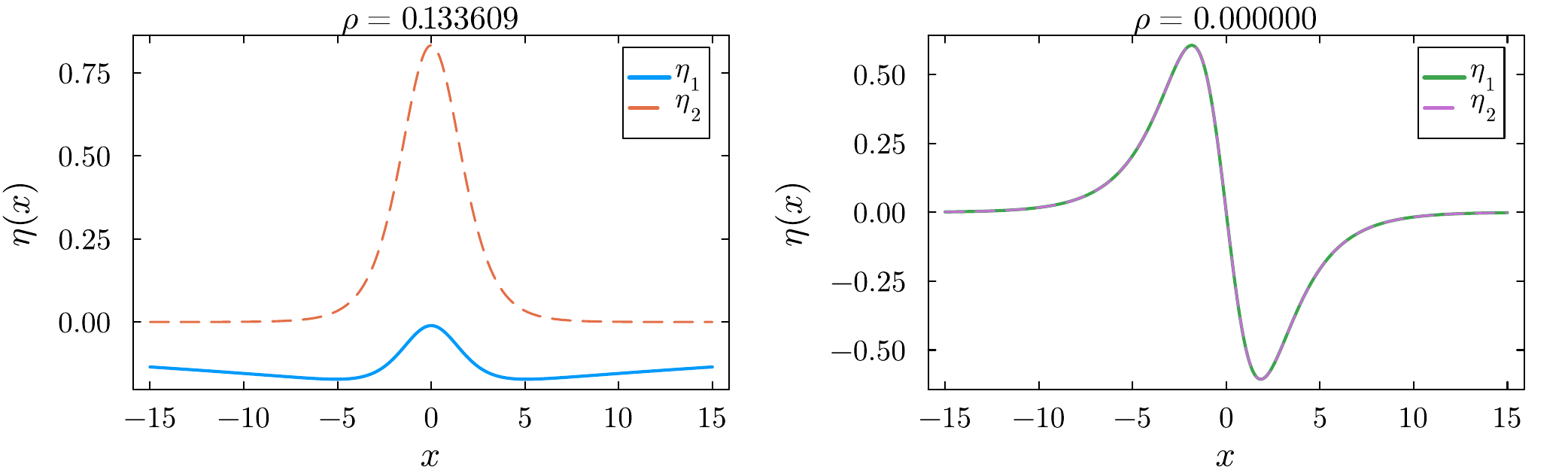}
    \caption{Bound modes for $\beta=\frac{1}{4}$ and $\omega=\frac{\sqrt{3}}{2}$.}\label{fig:bound_modes_Bowcock}
\end{figure}

\begin{figure}
    \centering
     \includegraphics[width=1\textwidth]{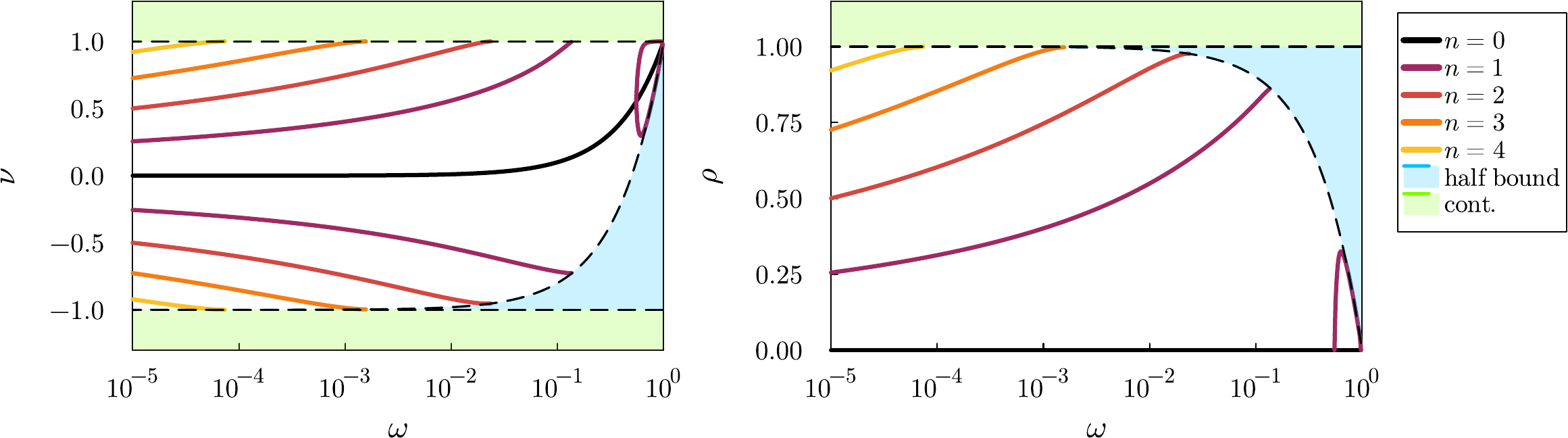}
     \caption{Eigenfrequencies of bound modes for $\beta=1/4$ as a function of $\omega$.}\label{fig:bound_modes_scan}
\end{figure}

\begin{figure}
    \centering
    \includegraphics[width=1\textwidth]{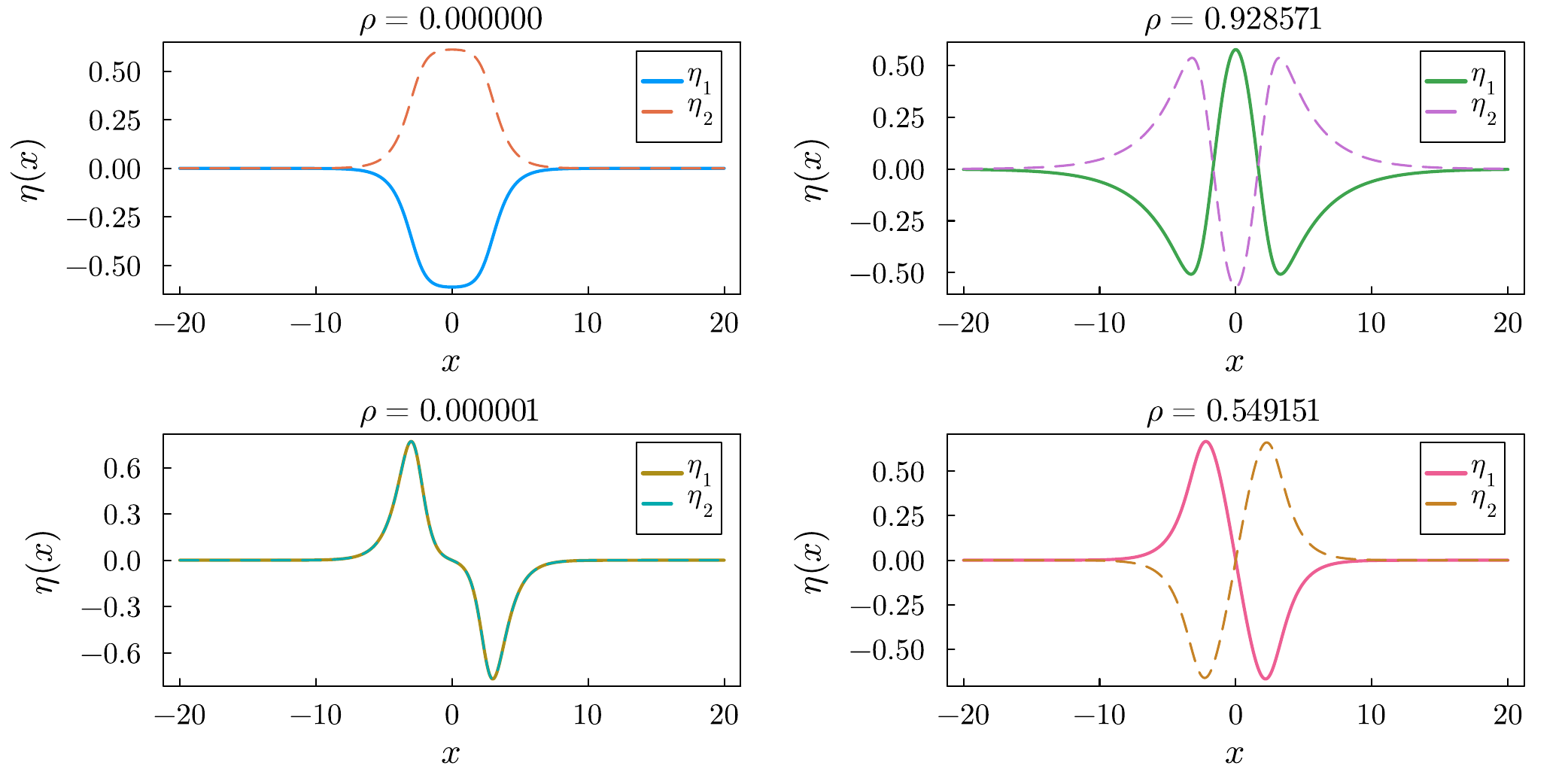}
    \caption{Bound modes of a small frequency Q-ball $\beta=1/4$, $\omega=0.01$. }\label{fig:bound_modes_small_omega}
\end{figure}

For $\beta=\frac{1}{4}+\epsilon$ the Q-ball frequency can be small and the Q-ball profile looks like two weakly bound kinks. In the spectrum there are two modes that can be interpreted as symmetric and antisymmetric combinations of translational modes of the kinks. The symmetric combination, which has  frequency $\rho=0$, is the translational mode of the Q-ball. The antisymmetric combination, with $\rho>0$, represents the oscillation of the width of the Q-ball. Moreover, as seen on the right-hand side of figure \ref{fig:bound_modes_scan}, more modes  appear as $\omega\to 0$, since the potential generated by the Q-ball has a wide well, which can support many modes. An example of such a case is shown in figure~\ref{fig:bound_modes_small_omega}. 

\subsection{Mode decay}\label{sec:decay}
Considering perturbations of the Q-ball
we repeat the same procedure as in \cite{Bowcock}, i.e.\ we examine a one-parameter squashing/stretching  perturbation of the form
\be
\phi \to \phi_\lambda=\sqrt{\lambda}e^{i\omega t}f(\lambda x)\, ,
\ee
where $\lambda$ is a positive parameter.
Note that these deformations do not affect the charge of the Q-ball. Hereafter we consider the most interesting case of a twofold degenerate vacuum
and take $\beta=\frac{1}{4}$, $\lambda=1.05$ and $\omega=\frac{\sqrt{3}}{2}$.
The decay of the perturbation is shown in Fig.~\ref{fig:mode_decay}.
In the Appendix we present and comment on a similar plot for the $\beta=0$ case studied in~\cite{Bowcock}.
\begin{figure}
    \centering
    \includegraphics[width=0.75\textwidth]{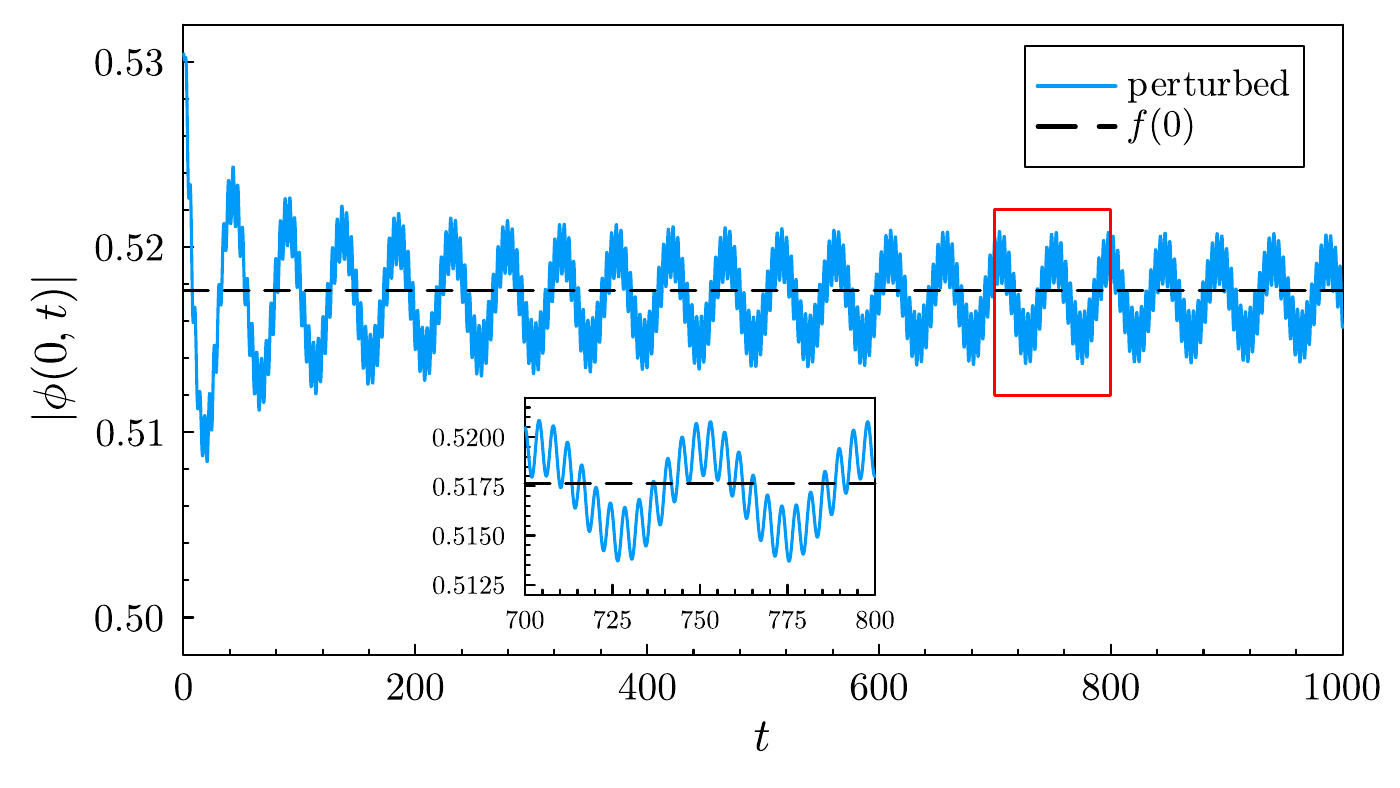}
    \caption{Evolution of the amplitude of the perturbed Q-ball $|\phi(0,t)|$  at the centre of the configuration for $\beta=\frac{1}{4}$, $\omega=\frac{\sqrt{3}}{2}$ and $\lambda=1.05$. }\label{fig:mode_decay}
\end{figure}
Considering the longer time evolution of the initially squashed Q-ball, we see that the perturbation decays with time, although the decay rate slows down. However, there is a pattern, which is similar to the Manton-Merabet law  \cite{Manton_1997} for the decay of the shape mode of the $\phi^4$ kink:  $A(t)\sim t^{-1/2}$.

\begin{figure}
    \centering
    \includegraphics[width=1\textwidth]{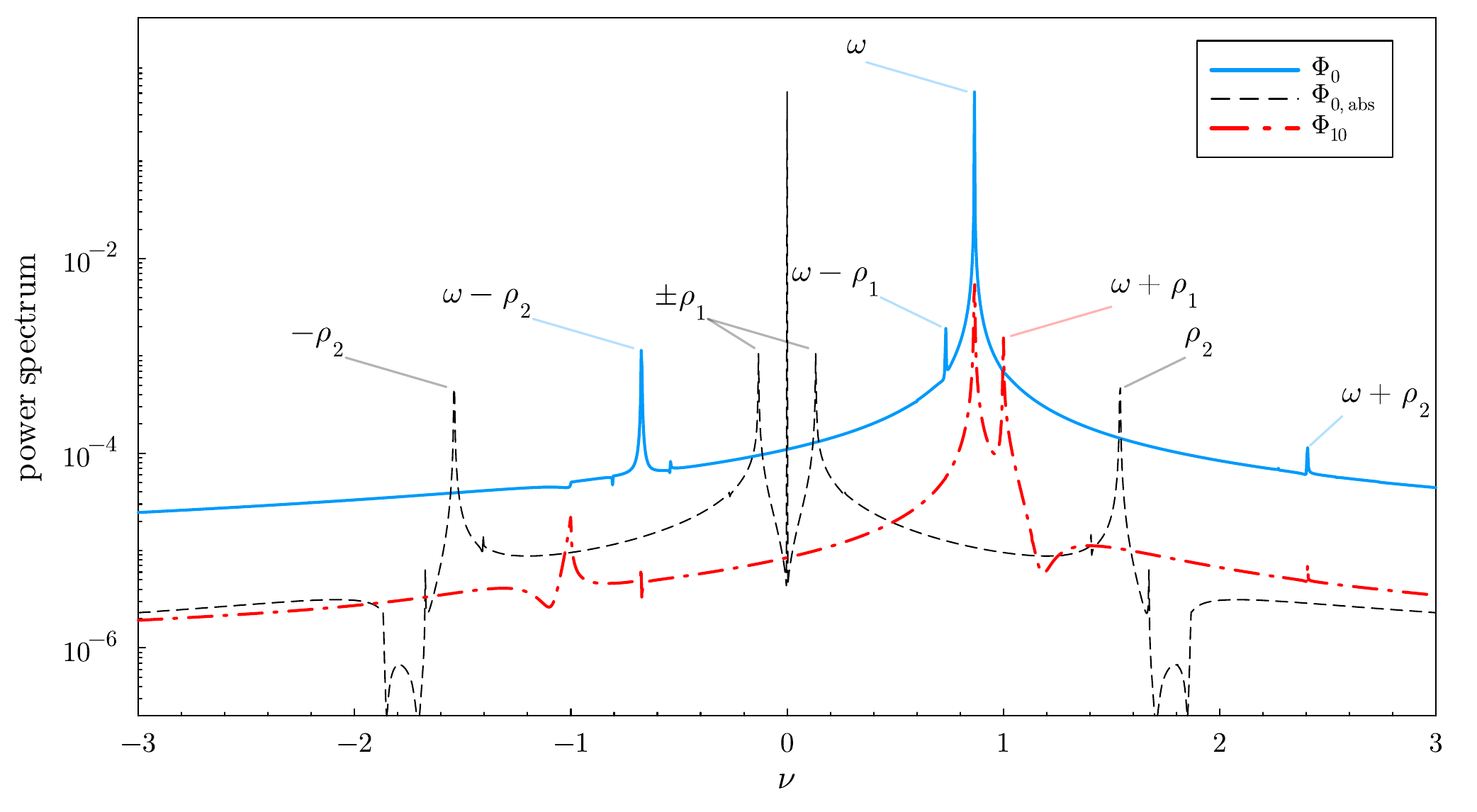}
    \caption{Power spectra of the fluctuations of the field $\phi$ (blue solid line) and of its absolute value $|\phi|$ (black dashed line) at the origin $x=0$, and of the field $\phi$ at $x=10$ (dash-dotted red line), all with $\beta=\frac14$.} \label{fig:power_spec_paul}
\end{figure}

In Figure~\ref{fig:power_spec_paul} we show the power spectra of the field $\phi$ at the centre and at $x=10$, and, following \cite{Bowcock}, the power spectrum of the magnitude of the field $|\phi(0,t)|$. Both the spectra of the field values and of their absolute values carry valuable information. Below we examine explicitly what are the differences between these two approaches.

Let us suppose that the field can be represented as a sum of modes oscillating with frequencies  $\nu_n$:
\begin{equation}
    \phi(0,t)\approx\sum_n c_n e^{i\nu_n t}.
\end{equation}
This yields
\begin{equation}
    |\phi(0,t)|^2=\phi\phi^*\approx\sum_{n,m} c_nc_m^* e^{i(\nu_n-\nu_m) t} \, .
\end{equation}
Thus, there are configurations with the same absolute value of the field $|\phi(0,t)|$ but different frequencies.

First, we consider the features of the power spectrum of perturbations at the center of the Q-ball, see the blue curve in Figure ~\ref{fig:power_spec_paul}.
The most prominent peaks in the spectrum correspond to the frequencies (in ascending order) $-0.673$, $0.734$, $0.866$ and $2.405$ .\\
\begin{itemize}
\item Clearly, the highest peak at $\nu=0.866$ corresponds to the frequency of the stationary Q-ball, $\omega =\sqrt{3}/2$. 
\item The peak at $\nu=0.734$ corresponds to the oscillational  mode with $\nu=\omega-\rho_{1}$ with $\rho_1=0.1336$.
\item At $\nu=\omega+\rho_1=0.9996<1$ there is a peak (barely visible at $x=0$ but quite prominent at $x=10$ - red dashed line) corresponding to the second frequency of the mode. Although $\nu=\omega+\rho_1$ is very close to 1 it is easy to verify numerically that it is still below the threshold, see figure \ref{fig:bound_modes_Bowcock}. 
\item The peak at $\nu=-0.673$ corresponds to the lower real part of the quasinormal mode with the real part of the frequency $\rho_2=1.539$. 
\item The second frequency of the quasinormal mode, corresponding to the propagating component, is visible at $\nu=\omega+\rho_2=2.405$. 
\end{itemize}

The mode $\omega-\rho_1 $ is bounded while the mode $\omega-\rho_2$ is half-propagating. Indeed, considering the power spectrum of fluctuations at $x=10$, see the red dash-dotted curve in Figure ~\ref{fig:power_spec_paul}, we can clearly identify the Q-ball frequency $\omega$ and threshold frequencies $\pm  1$. Also, small peaks at $\omega \pm \rho_2$ are visible there.

The power spectrum of fluctuations of absolute value of the field at the origin is displayed in Figure ~\ref{fig:power_spec_paul}, black curve. The most prominent peak is at at $\omega=0$ which corresponds to the net contribution of all frequencies,  $\omega_n-\omega_n$=0. Another peak, at $\pm 0.132$, corresponds to $\pm(\omega_3-\omega_2)$ and the peaks at $\pm 1.541$ correspond to $\pm(\omega_3-\omega_1)$.

\section{Radiation pressure}\label{sec:radiationpressure}
In this section, we investigate how Q-balls move when exposed to scalar radiation. We expect that for small amplitudes (much smaller than the amplitude of the Q-ball) the waves would scatter according to linearized equations. The scattered waves can have different momenta than the initial wave, and the excess of the momentum carried by waves can be related to the force acting on the Q-ball. In particular, we will show that the radiation can both push (positive radiation pressure, PRP) or pull (negative radiation pressure, NRP) the Q-ball depending on the composition of the incoming wave.

Following \cite{NRP_IN_BE,Forgacs:2008az},
let us consider a field within a finite interval $x\in [-L,L]$, where $L\gg1$, containing a Q-ball and a wave scattered off the Q-ball. The total momentum of the field in this interval is given by
\begin{equation}
P=\int_{-L}^{L}\mathcal{P}\,dx=-\int_{-L}^{L}T_{tx}\,dx=-\int_{-L}^{L}(\phi_t\phi_x^*+\phi_t^*\phi_x)dx\,,
\end{equation}
where $\mathcal{P}$ is the momentum density, and $T$ is the energy-momentum tensor.
A moving Q-ball then has momentum $P=E\gamma v$. A wave scattered off the Q-ball carries momentum which can be obtained from the conservation law  $\partial_\mu T^{\mu\nu}=0$ 
\begin{equation}
\partial_t\mathcal{P}=-\partial_xT_{xx}=-\partial_x\left[\partial_t\phi\partial_t\phi^*+ \partial_x\phi\partial_x\phi^*-V\left(|\phi|^2\right)\right].
\end{equation}
Integrating the left-hand side of the above expression and averaging over a period gives the rate of change of the momentum within the interval. The right hand is a total derivative, so integration gives only boundary terms. Assuming that the wave asymptotically (for $x\approx \pm L$) has the form  $\phi=A(x)e^{i(\nu t-kx)}$, where $A(x)$ is a slowly changing function, we find that the rate of change of the momentum inside of the interval is equal to the flux of momentum through the boundaries of the interval:
\begin{equation}
    \frac{dP}{dt} = \int_{-L}^L \partial_t\mathcal{P} dx = 2k^2\left[A^2(-L)-A^2(L)\right].
\end{equation}
Assuming that the amplitudes of the waves are small compared to that of the Q-ball, we can identify $\frac{dP}{dt}$ as a force $F$ acting on the Q-ball.

\begin{figure}
    \centering
    \includegraphics[width=0.49\textwidth]{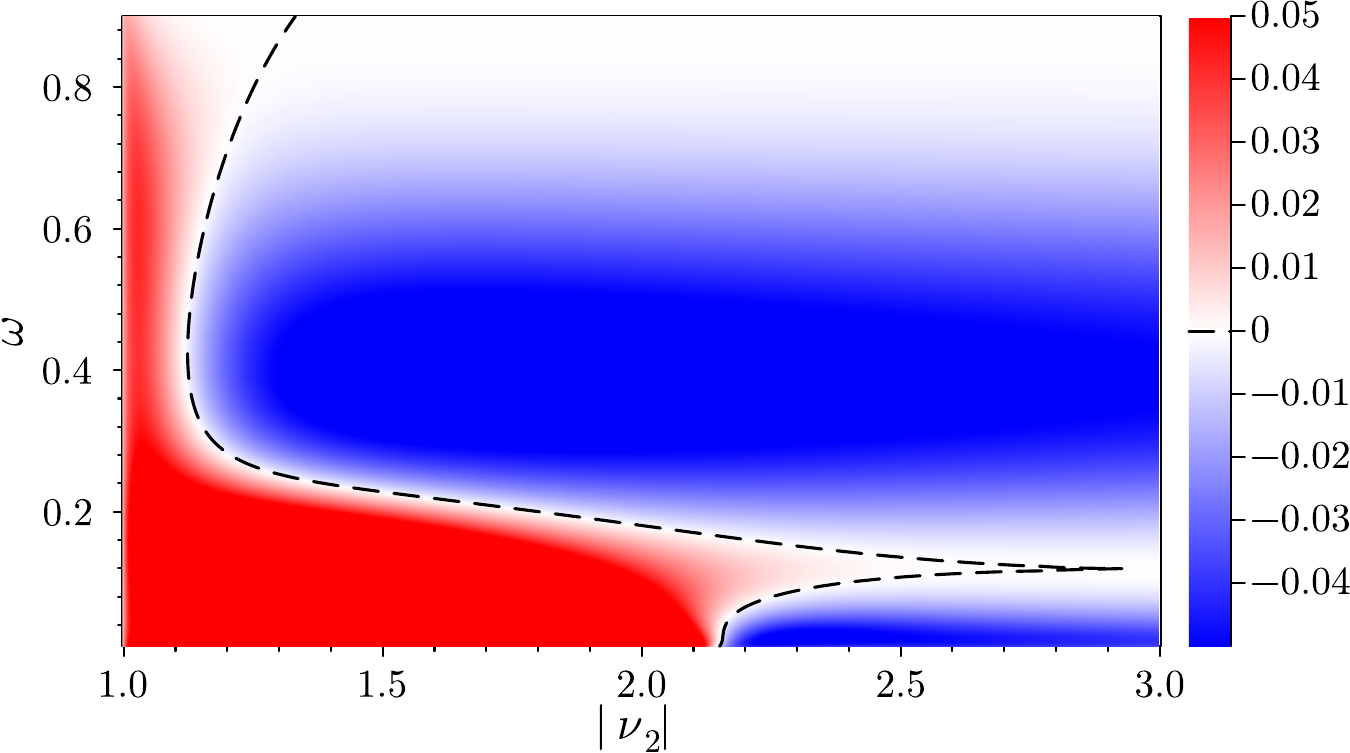}
    \includegraphics[width=0.49\textwidth]{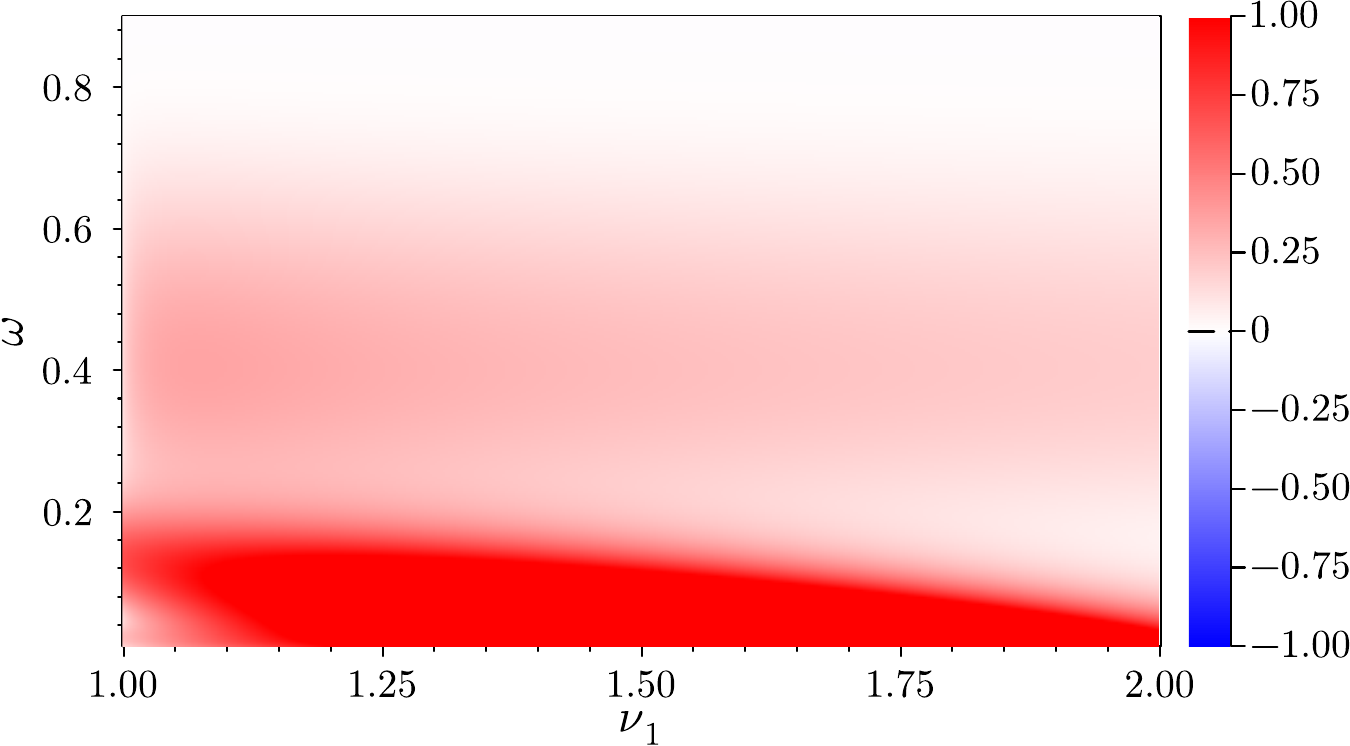}
    \caption{Force (divided by $A^2$) exerted on Q-balls for $\beta=\frac14$ as a function of $\nu_2$ and $\omega$ in the case of incident wave in the second sector.}\label{fig:radiation_pressure_force}
\end{figure}

\begin{figure}
    \centering
    \includegraphics[width=1\textwidth]{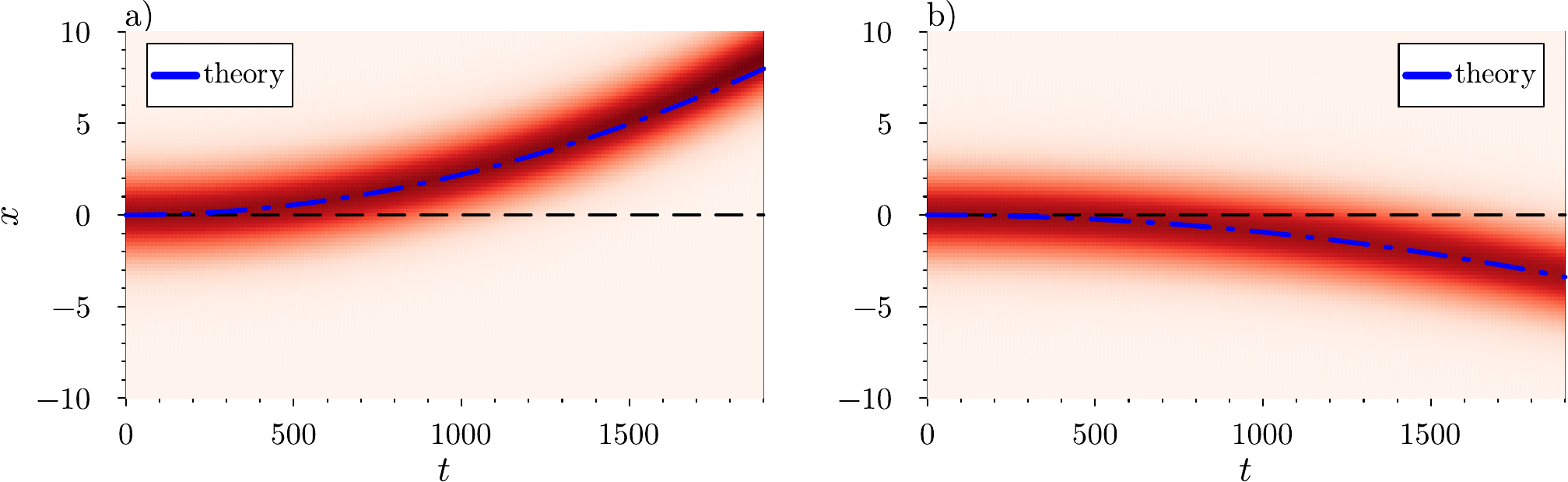}
    \caption{Motion of Q-balls under the influence of a wave in a) first channel and b) second channel. $\beta=0.26$, $\omega=0.6$, $\rho=2.1$,  $A=0.01$. Colour intensity indicates the amplitude of the field.}\label{fig:radiation_pressure}
\end{figure}

\begin{figure}
    \centering
     \includegraphics[width=1\textwidth]{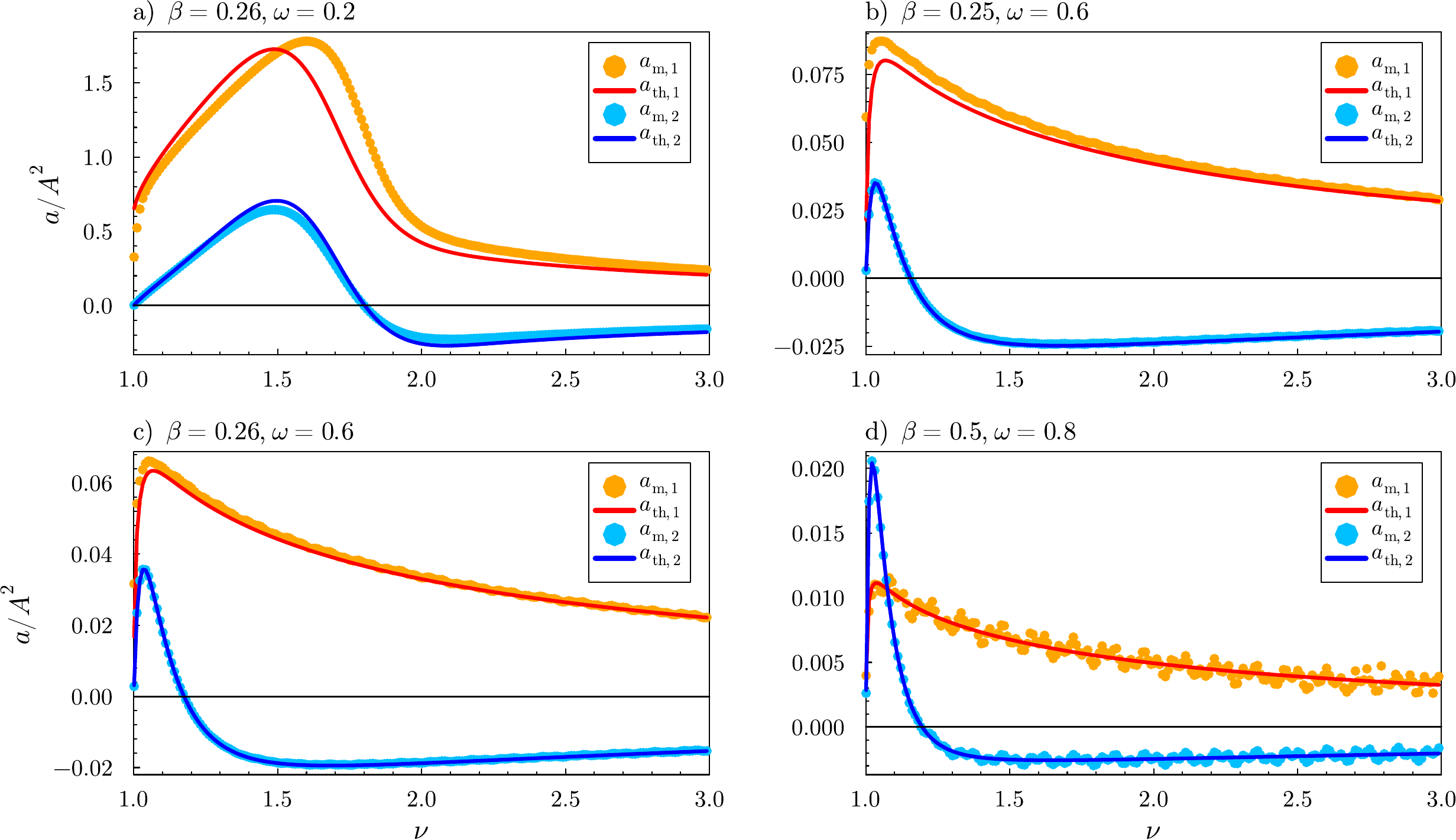}
     \caption{Comparison of measured acceleration (dots) with theoretical prediction (lines).}\label{fig:comparison_NRP}
\end{figure}

A single mode consists of two components with frequencies $\nu_1=\omega+\rho$ and $\nu_2=\omega-\rho$. Therefore, the asymptotic form of the field for which the incident wave propagates in channel
 $j$ is
\begin{equation}
    \phi(x\to-\infty)=A\sum_{l=1,2}\left(\delta_{lj}+r_j\right)e^{i(\nu_lt+k_lx)}
\end{equation}
\begin{equation}
    \phi(x\to\infty)=A\sum_{l=1,2} t_l e^{i(\nu_lt-k_lx)}
\end{equation}
The rate of momentum transfer to the Q-ball is the sum of contributions coming from all scattered waves:
\begin{equation}
\label{effective_force}
    F_j = 2A^2\sum_{l=1,2}k_l^2(\delta_{lj}+R_l-T_l)
\end{equation}
where $R_l=|r_l|^2$ and $T_l=|t_l|^2$ are reflection and transmission coefficients in each channel and $k_l$ are wave numbers. Negative radiation pressure can occur if the transmission coefficient from a smaller to larger wave number channel is large enough. The only such scenario is when the incident wave corresponds to a frequency $\nu_2=\omega-\rho<-1$. The other frequency is equal to $\nu_1=2\omega-\nu_2$, so $|k_2|<|k_1|$. Figure \ref{fig:radiation_pressure_force} shows the force calculated from the solution of the linearized equation exactly in such a case. For values of $|\nu_2|$ close to the threshold 
positive radiation pressure is expected, while for larger $|\nu_2|$ there are regions where negative radiation pressure should be visible.
The red colour represents PRP and the blue, NRP. 
In the other case, with an incident wave in the first component, the radiation pressure was always positive.
An example of both positive and negative radiation pressure for $\beta=0.26$ and $\omega=0.4$ is shown in figure~\ref{fig:radiation_pressure}. The accelerations predicted from the effective force (\ref{effective_force}) agree well with the accelerations computed from numerical simulations of the full PDE, as presented in figure \ref{fig:comparison_NRP}.

In the same way as we calculated the force from the energy and momentum conservation law, we can calculate the rate of change of the charge of a Q-ball by integrating (\ref{charge})
\begin{equation}
    \frac{dQ}{dt}=\int dx\, \partial_x j_x= 2A^2\sum_{l=1,2}|k_l|(\delta_{lj}-R_l-T_l)
\end{equation}
Numerically we found this value to be of order $10^{-15}$, which means that to linear order the wave does not change the charge of the Q-ball. The result is not surprising because it follows from the unitarity of the $S$ matrix. 

\section{Conclusions}\label{sec:conclusions}
In this paper we have discussed how external perturbations affect Q-balls in (1+1) dimensions. Two examples are considered: (i) squashing perturbations of the Q-ball and (ii) interaction with incoming radiation. 
Considering the power spectrum of linearized perturbations about the Q-ball we find that there are both translational and phase shifting zero modes, as well as bounded, half-propagating and quasi-normal modes. We have shown that, depending on the nature of the incoming wave, the Q-ball can be pulled towards the source of the incident radiation, or it can be pushed away from it.

An interesting extension of this work would be to study perturbative excitations of  Q-balls in (3+1) dimensions. Another interesting question, which we hope to be addressing in the near future, is to investigate the decays of Q-balls caused by external perturbations.

We note, finally, that stationary rotating Q-balls can induce superradiant amplification for scattered outgoing waves \cite{Saffin}. An important assumption in \cite{Saffin} is that the incoming wave does not excite the translational mode of the Q-ball. As a result its position does not change during the process of  superradiant emission of radiation, related to the resonance energy transfer from the stationary Q-ball to the scattered wave. However, since, as we have seen, radiation pressure on the Q-ball pushes it to move, one may wonder whether this will affect the criteria for superradiance to occur, and it would be interesting to investigate this question further.

\section*{Acknowledgements}
PED was supported in part by the STFC under consolidated grant ST/T000708/1 “Particles, Fields and Spacetime”, and would like to thank the African Institute for Mathematical Sciences, South Africa, for hospitality during the last stages of this work.
TR was supported by the Polish National Science Center, 
grant NCN 2019/35/B/ST2/00059. TR and DC were supported by the Priority Research Area under the
program Excellence Initiative—Research University at
the Jagiellonian University in Kraków.
YS would like to thank the
Hanse-Wissenschaftskolleg Delmenhorst for support. 
We would like to thank Mikhail Smolyakov for pointing out the presence of the Lorentz mode, which was missed in the previous version of the manuscript.

\appendix 
\section{Metastable Q-balls}\label{metaqballs}
In the main part of this paper we studied the behavior of Q-balls for models with $\beta\geq\frac{1}{4}$, when potentials have one ($\phi=0$ for $\beta>\frac{1}{4}$) or two vacua ($|\phi|=0, \sqrt{2}$ for $\beta=\frac{1}{4}$). For $\beta<\frac{1}{4}$ the minimum at $\phi=0$ is just a local minimum, and therefore it is a false vacuum. However, certain restricted configurations, if not excited too much, can exist around the false vacuum, at least in the classical theory.
In \cite{Bowcock} the decay of such a configuration for the model with $\beta=0$ was discussed. We have repeated the same analysis as in section \ref{sec:decay} but for $\beta=0$. In this appendix we report on these results. 

\begin{figure}
    \centering
     \includegraphics[width=0.75\textwidth]{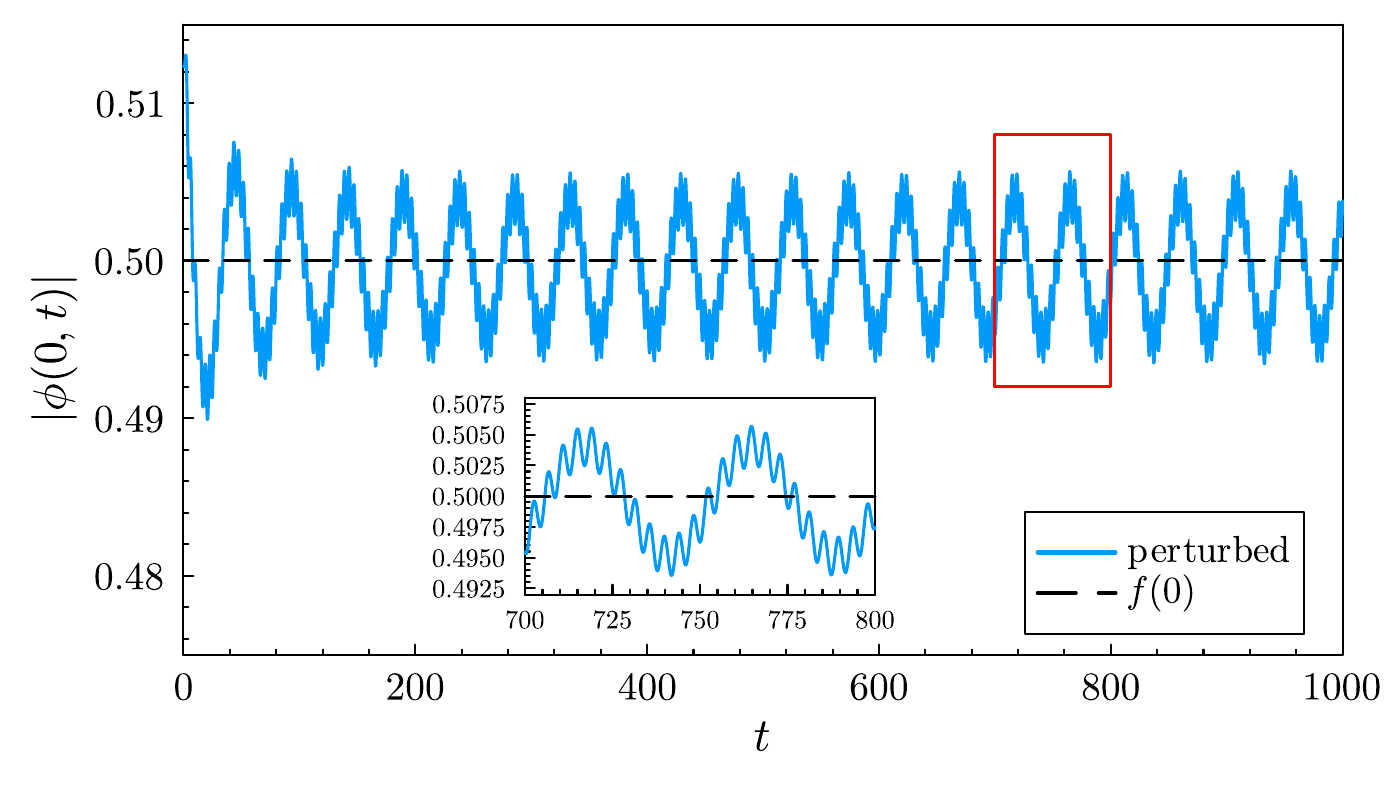}
     \caption{Evolution of the amplitude of the perturbed Q-ball $|\phi(0,t)|$  at the centre of the configuration for $\beta=0$, $\omega=\frac{\sqrt{3}}{2}$ and $\lambda=1.05$. }\label{fig:mode_beta_decay}
\end{figure}

\begin{figure}
    \centering
    \includegraphics[width=1\textwidth]{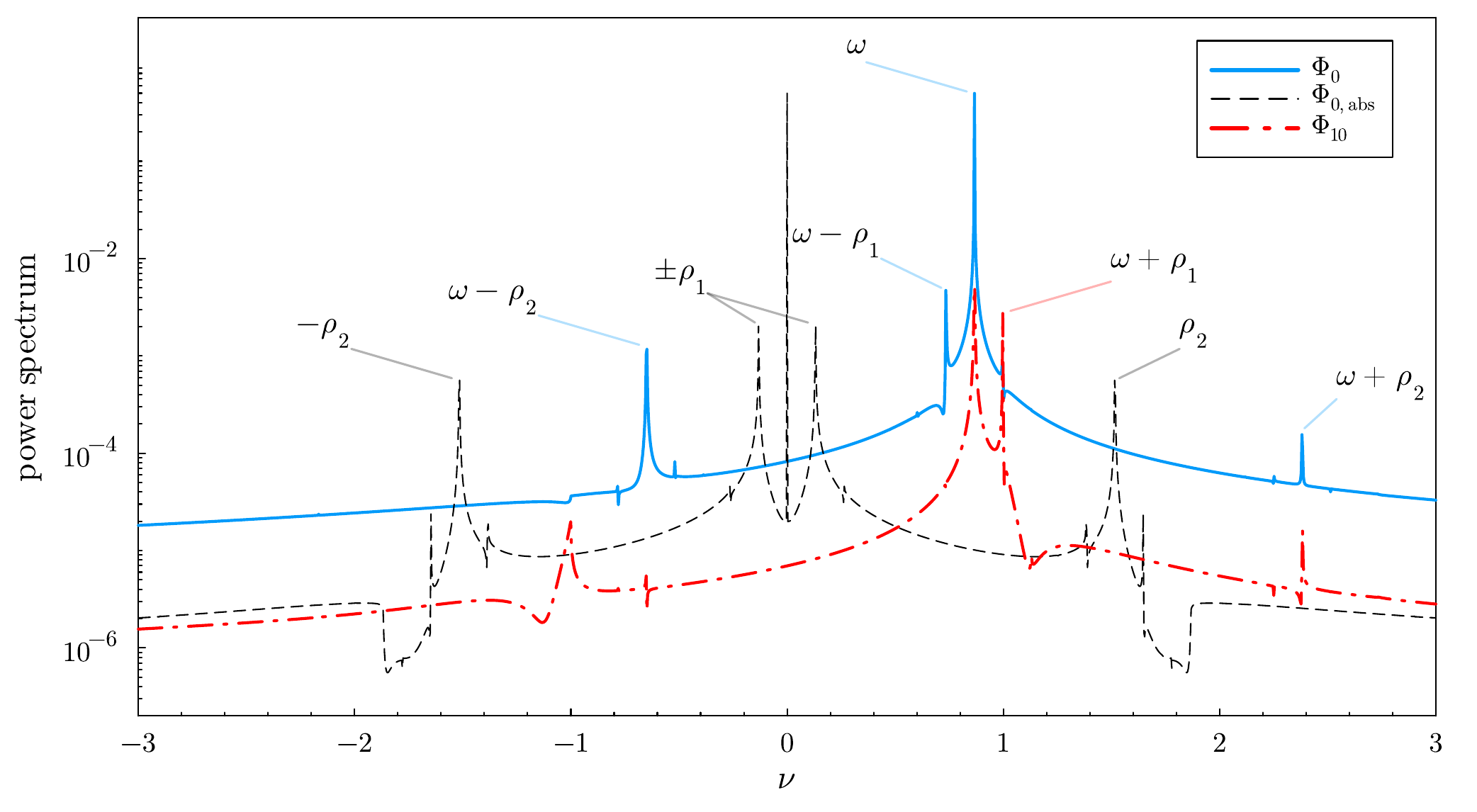}
    \caption{Power spectra of the fluctuations of the field $\phi$ (blue solid line) and of its absolute value $|\phi|$ (black dashed line) at the origin $x=0$, and of the field $\phi$ at $x=10$ (dash-dotted red line), all with $\beta=0$.} 
 \label{fig:power_spec_paul_beta}
\end{figure}

We present the decay in figure \ref{fig:mode_beta_decay}, while the power spectra are presented in figure \ref{fig:power_spec_paul_beta}. A direct comparison with figure \ref{fig:power_spec_paul} shows a very similar spectral structure with a single bound and one quasi-normal mode.  The bound mode is for $\rho_1=0.1317$ and the QNM is for 
$\rho_2=1.5150692 + 9.96\cdot10^{-5}i$. Measured peaks: -0.650, 0.731, 0.997, 2.383 are consistent with the bound mode \mbox{$\omega+\rho_1=0.9977$}, $\omega-\rho_1=0.7343$ and the QNM 
$\omega+\mathop{\textrm{Re}}\rho_2=2.3811$ and 
\mbox{$\omega-\mathop{\textrm{Re}}\rho_2=-0.64904$}.
In \cite{Bowcock} the authors claim that the persisted oscillations are due to the bound state of two Q-balls, some form of a breather. However, since the frequencies match so well with the bound and the quasinormal mode, we believe that their claim is overstated.

\normalem

\bibliographystyle{JHEP}
\bibliography{Qballs_paper}
\end{document}